\documentclass[12pt,preprint]{aastex}

\shorttitle{Chlorine in the ISM}
\shortauthors{Sonnentrucker et al.}

\begin{document}


\title{Chlorine in the Galactic ISM: Revised $f$-values with {\it FUSE} and STIS\altaffilmark{1}}


\author{P. Sonnentrucker\altaffilmark{2}, S. D. Friedman\altaffilmark{3} and D. G. York\altaffilmark{4}}


\altaffiltext{1}{Based on observations made with the NASA/ESA {\it Hubble Space 
Telescope}, obtained from the Data Archive at the Space Telescope Science 
Institute, which is operated by the Association of Universities for Research 
in Astronomy, Inc., under NASA contract NAS 5-26555.}
\altaffiltext{2}{Department of Physics and Astronomy, Johns Hopkins University,
 3400 North Charles Street, Baltimore, MD 21218}
\altaffiltext{3}{Space Telescope Science Institute, 3700 San Martin Dr., Baltimore, MD 21218}
\altaffiltext{4}{Department of Astronomy and Astrophysics, University of 
Chicago, 5640 South Ellis Avenue, Chicago, IL 60637.}


\begin{abstract}

\ion{Cl}{1} is the atomic species most directly coupled to molecular hydrogen due to its chemistry. Its weakest lines are thereby probably the best tracer of optically thick H$_2$ components in diffuse clouds. We report on the empirical determination of the oscillator strengths for four \ion{Cl}{1} absorption lines predicted to be weak and often detected toward moderately reddened sight lines observed with the {\it Far Ultraviolet Spectroscopic Explorer} ({\it FUSE}). We compared our oscillator strength estimates with the oscillator strength calculations listed in Morton (2003). We find that our empirical oscillator strength values for the \ion{Cl}{1} 1004, 1079, 1090 and 1094 \AA\ lines differ from the theoretical predictions by factors of $\sim$3.1, 1.2, 2.4 and 0.42, respectively. We briefly discuss the value of \ion{Cl}{1} as tracer of molecular gas for our star sample.

\end{abstract}





\section{Introduction}

Due to its unique chemistry, chlorine plays an important role in characterizing the neutral
gas components (Jura 1974; Dalgarno et al. 1974) in the interstellar medium (ISM). In diffuse clouds, the detection of chlorine in its various ionization states allows us to disentangle the diffuse molecular gas components from the neutral atomic gas components (Jura \& York 1978) along sight lines exhibiting complex structures that are often unresolved with the current instrumentation (e.g., Rachford et al. 2002). Most importantly, \ion{Cl}{1} offers an alternate and more reliable way of tracing cold H$_2$ gas components in diffuse clouds since, unlike CO, the chlorine abundance remains unaffected by molecular processes such as chemical fractionation and photodissociation (Sonnentrucker et al. 2002). 

\ion{Cl}{2} exhibits transitions solely below 1080 \AA, all with laboratory-determined $f$-values (Schectman et al. 1993, 2005). \ion{Cl}{1}, on the other hand, is predicted to have over 50 transitions between $\lambda$920 and $\lambda$1390 \AA\  (Morton 2003 and references therein). To date, only three lines, $\lambda$$\lambda$1088, 1097 and 1347 \AA, possess experimentally-determined $f$-values. Because of its strength the \ion{Cl}{1} 1347 \AA\ line is almost always optically thick and its use requires careful treatment of saturation effects (Jura \& York 1978; Jenkins et al. 1986; Harris \& Bromage 1984). The $\lambda$1088 \AA\ line often suffers from severe blending with an adjacent CO band (Federman 1986) while the $\lambda$1097 \AA\ line, due to its intrinsic weakness, requires significant reddening and high S/N ratio to be detectable along a given sight line. Consequently, accurate determination of the \ion{Cl}{1} column density and the chlorine cosmic abundance remained difficult with prior space-based missions such as {\it Copernicus} and {\it IUE} (Keenan \& Dufton 1990).

The {\it FUSE} wavelength range and especially its greater sensitivity -- compared to previous missions-- allowed us to revisit the role that chlorine can play in increasing our understanding of the structure in the diffuse atomic and diffuse molecular ISM. Since a reliable atomic database is necessary for such an endeavor, we proposed to empirically determine the oscillator strength ($f$-value) for 4 \ion{Cl}{1} lines particularly well suited for our study. The \ion{Cl}{1} 1004, 1079, 1090, 1094 \AA\ lines are predicted to be weak, are often detected toward low-to-moderately reddened stars and do not significantly suffer from blends with other species. We present the results of our $f$-value investigation for these 4 \ion{Cl}{1} lines toward the moderately reddened stars, HD91597, HD220057 and HD108 and three additional stars from the {\it FUSE} archive. Section 2 summarizes our observations and data reduction technique. Section 3 describes our analysis, compares our empirical $f$-value estimates to the theoretical calculations listed in Morton (2003), and discusses the value of \ion{Cl}{1} as a tracer of H$_2$.

\section{Observations and data reduction}

Three stars were successfully observed for our program (D064), HD91597, HD220057 and HD108. Their selection was based on the following criteria. 1) Existence of archival STIS data covering the \ion{Cl}{1} 1347 \AA\ range; 2) simple velocity structure to ensure proper correction for saturation, when present and; 3) small or negligible amounts of CO to minimize blending with the \ion{Cl}{1} 1088 \AA\ line.

The {\it FUSE} data were processed with version 3.0.8 of the CalFUSE pipeline. For each sight line, the individual spectra were co-added by cross-correlation with respect to the brightest exposure, in each segment. The FP-split technique was used while acquiring observations toward HD91597 and HD220057 in order to reach S/N ratios $\ge$ 30 per resolution element, ratios calculated to obtain detection of the weak \ion{Cl}{1} 1097 \AA\ at a 3$\sigma$ level. While this level of detection was achieved toward HD108, the data collected toward HD91597 and HD220057 only allowed for a detection of the 1097 \AA\ line at a 1.8$\sigma$ and 2.6$\sigma$ level, respectively (Table 1). For all targets presented here, the S/N ratio ranges from 15 to 41. Figure~1  displays the co-added spectra obtained with the LiF1A and LiF2A segments, respectively.

The calibrated {\it HST} data were retrieved from the MAST archive. All stars were observed with the STIS instrument, using the 0.1$''$ $\times$ 0.03$''$ or 0.2$''$ $\times$ 0.2$''$ apertures and the E140H grating (P8484: Jenkins; P9434-P8241: Lauroesch). No additional reductions were performed on these datasets.





\section{Results}

In order to determine whether corrections need to be applied to the theoretical $f$-values listed in Morton (2003) for the  \ion{Cl}{1} 1004, 1079, 1090 and 1094 \AA\ transitions, one must first determine the total \ion{Cl}{1} column density toward each sight line. Since both \ion{Cl}{1} lines at 1347 and 1097 \AA\ have experimentally-determined $f$-values and because $f$(1347)/$f$(1097)$\sim$17.4 (Schectman et al. 1993), these two lines uniquely constrain both the linear and the flat part of the curve of growth for a given sight line, for the column densities observed here. We, subsequently, designed our observing strategy in order to obtain equivalent widths for both lines and we derived the corresponding \ion{Cl}{1} column density using a one-component curve of growth. We then measured the equivalent width of the 4 \ion{Cl}{1} lines under study and determined the best-fit $f$-values to the previously-constrained curves of growth with a least-squares technique.

As can be seen in Figs.~1 and 2, the \ion{Cl}{1} 1090 and  1079 \AA\ lines are intrinsically weak and are thereby only detected toward the most reddened stars in our sample. In order to both better constrain the oscillator strength of these 2 lines and to test the robustness of our revised $f$-values for the stronger lines ($\lambda$$\lambda$1004 and 1094 \AA), we measured the equivalent widths of all 6 \ion{Cl}{1} transitions toward three additional sight lines. The choice of these stars was driven by their reddening, similar to that of HD108, the existence of both archival {\it FUSE} and STIS data and the extensive knowledge regarding their respective sight line physical conditions (Rachford et al. 2002; Sonnentrucker et al. 2003; Pan et al. 2005), of importance to consider saturation effects. Table~1 lists our equivalent width measurements for all the \ion{Cl}{1} lines toward the program stars selected for this study (upper half) and toward the 3 additional moderately-reddened sight lines extracted from the {\it FUSE} and STIS archives (lower half). When measurements are available in two segments, we report the equivalent width weighted mean calculated by weighting each measurement by its corresponding error. All errors include fixed-pattern noise and continuum placement uncertainties and represent 1$\sigma$ formal errors alone. Figure 2 displays  the \ion{Cl}{1} 1097 \AA\ line (right panels) and the \ion{Cl}{1} 1079 \AA\ line (left panels) in the heliocentric velocity frame. 

Figure~3 (upper panels) shows the one-component curve of growth (COG) obtained from the \ion{Cl}{1} 1097 and 1347 \AA\  measurements (black symbols) toward our program stars HD91597, HD220057 and HD108. The red symbols show the best fit of the $\lambda$$\lambda$1004, 1079, 1090 and 1094 \AA\ line equivalent widths to each COG using our revised $f$-values. The lower panels display the measurements and respective COGs for the archival stars used to test our results. As can be readily seen from Table~2, our target selection covers a wide enough range in column density that each \ion{Cl}{1} line is on the plotted linear part of the COG toward at least 2 sight lines and suffers various degrees of saturation toward the rest of the sample, thus ensuring that our revisions to the theoretical $f$-values are not significantly affected by saturation. 

Other potential sources of error when revising the $f$-values empirically arise from blends between chlorine and other ISM species or blends with instrumental artifacts such as fixed-pattern noise or local sensitivity deficiencies. We determined that none of the lines investigated here suffered from such artifacts by measuring the equivalent widths in different segments. The \ion{Cl}{1} lines at 1004, 1079 and 1090 \AA\ are also all free of blends with other interstellar species. The \ion{Cl}{1} 1094 \AA\ is the only line under investigation that could potentially be affected by absorption caused by an intervening H$_2$ $J =$ 6 line ($\lambda$1094.80 \AA). To determine to what extent this H$_2$ line may affect our results, we estimated the H$_2$ column density using the $J=$ 6 absorption lines arising from all other H$_2$ ro-vibrational transitions present in the {\it FUSE} bandpass. Between 4 and 7 additional, unblended $J=$ 6 lines were measured. We determined that the equivalent width of this H$_2$ line is less than the error in the equivalent width of the \ion{Cl}{1} 1094 \AA\ line in all cases. We are thereby confident that our $f$-value estimate of the \ion{Cl}{1} 1094 \AA\ line is not significantly affected by H$_2$ blending in our star sample.

Table~3 lists the wavelengths of the 4 \ion{Cl}{1} lines we investigated and compares the theoretical $f$-values listed in Morton (2003; and references therein) with our proposed recommendations. The last column indicates whether the chlorine lines were subject to potential blends. Within our measurement errors, we find that the ratios of the theoretical $f$-values to our empirical $f$-values differ by factors $\sim$3.1 ($\lambda$1004 \AA), $\sim$1.2 ($\lambda$1079 \AA), $\sim$2.4 ($\lambda$1090 \AA) and $\sim$0.42 ($\lambda$1094 \AA). 

Our analysis also shows that \ion{Cl}{1} can sometimes be used to draw additional conclusions regarding the structure of the molecular gas toward a given sight line. Because \ion{Cl}{1} results from rapid ion-molecule reaction between Cl$^+$ and H$_2$ once molecular hydrogen is optically thick, \ion{Cl}{1} constitutes a good tracer for molecular gas components in general. Hence, one can use the \ion{Cl}{1} gas structure to infer the distribution of H$_2$ wherever it is optically thick and thereby often difficult to measure accurately. For the star sample presented here our chlorine analysis indicates that $b$-values of the order of those listed in Table~2 are most appropriate to constrain the H$_2$ column densities arising from the lowest rotational levels if saturation is observed (e.g., Sonnentrucker et al. 2003). Adopting the \ion{H}{1} column densities from the literature (see Table~2), we estimated the \ion{Cl}{1} to \ion{H}{1} column density ratios. Our results further indicate that, for these stars, the variation in this ratio by almost a factor 15 is mostly due to variations in the \ion{Cl}{1} column density which, in turn, is mostly tracing the increasing fraction of optically thick molecular hydrogen present along each sight line (Jura \& York 1978).

\acknowledgments

P.S. acknowledges support of grant NAG5-13698. We thank the referee for very helpful comments.

\clearpage

\begin{figure}
\epsscale{0.6}
\plotone{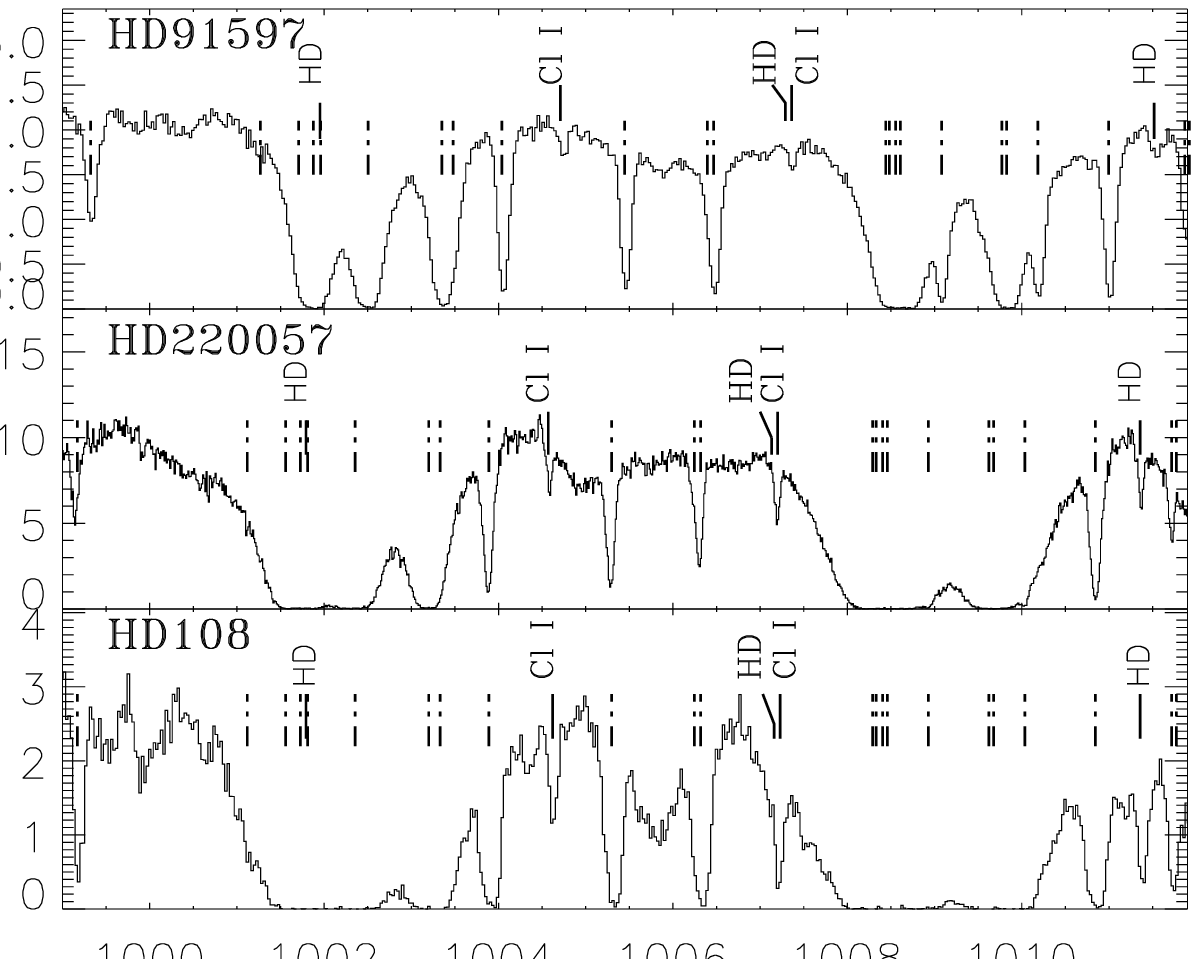}
\plotone{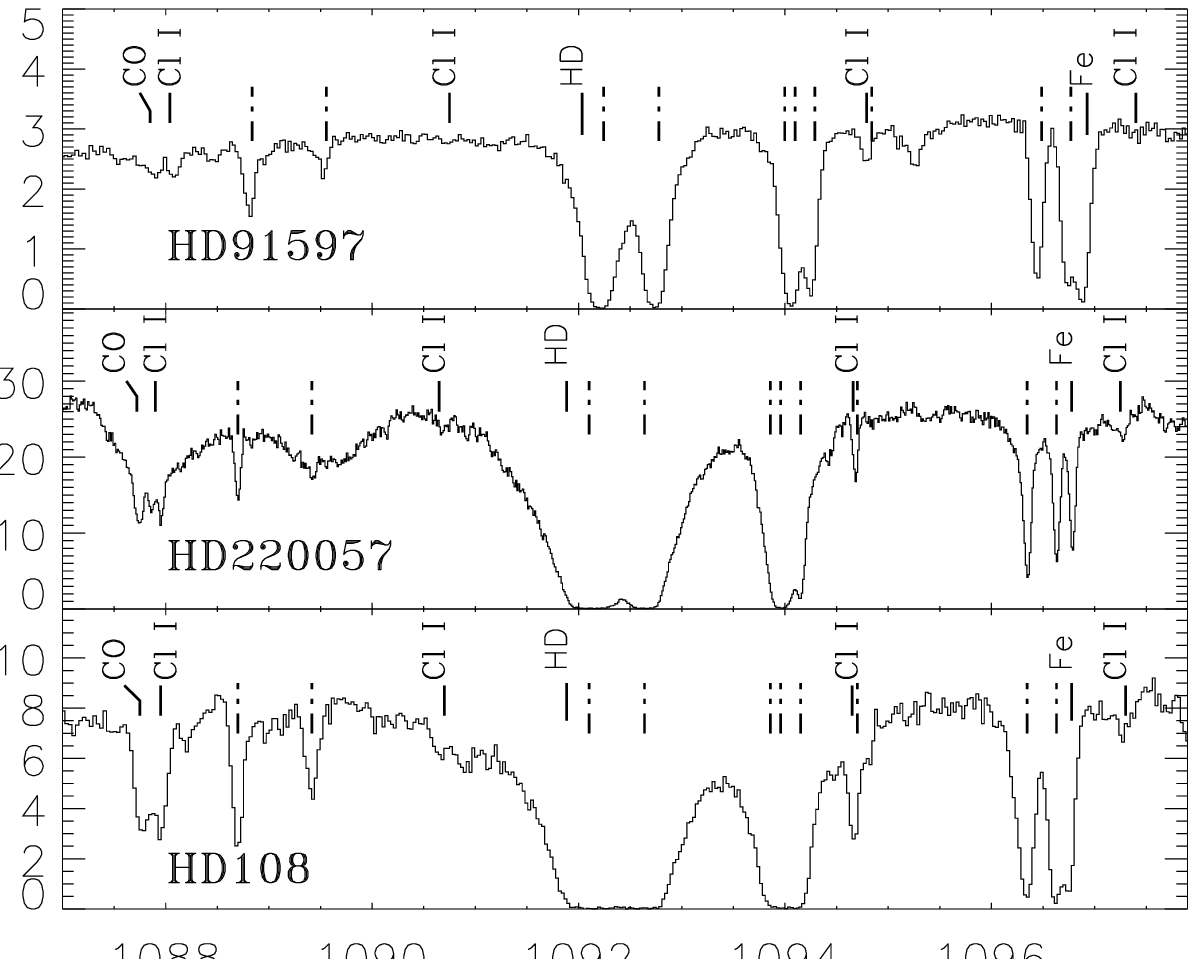}
\caption{Spectra of the \ion{Cl}{1} 1004, 1088, 1090, 1094 and 1097 \AA\ lines observed with {\it FUSE} toward HD91597, HD220057 and HD108 in the wavelength region covered by LiF1A and LiF2A. Dash-dotted tickmarks indicate the position of the H$_2$ lines. Solid tickmarks show the position of \ion{Cl}{1} and a few intervening HD, CO and Fe lines.\label{fig1}}
\end{figure}

\clearpage

\begin{figure}
\epsscale{0.75}
\plotone{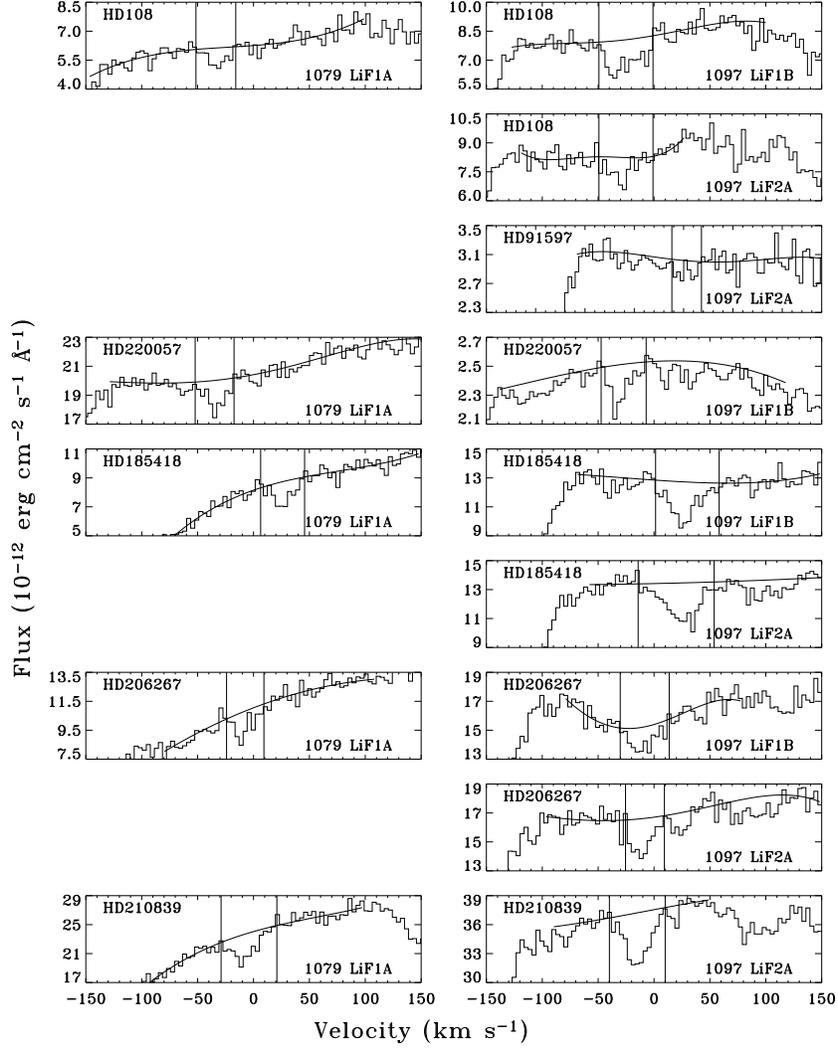}
\caption{Spectra and continuum fits (smooth curves) of the \ion{Cl}{1} 1079 \AA\ (left panels) and \ion{Cl}{1} 1097 \AA\ lines (right panels) for the stars and segments where the lines are detected. The velocity scale is heliocentric. The vertical lines indicate the equivalent width intervals of integration for each line. \label{fig2}}
\end{figure}

\begin{figure}
\epsscale{0.75}
\rotatebox{90}{\plotone{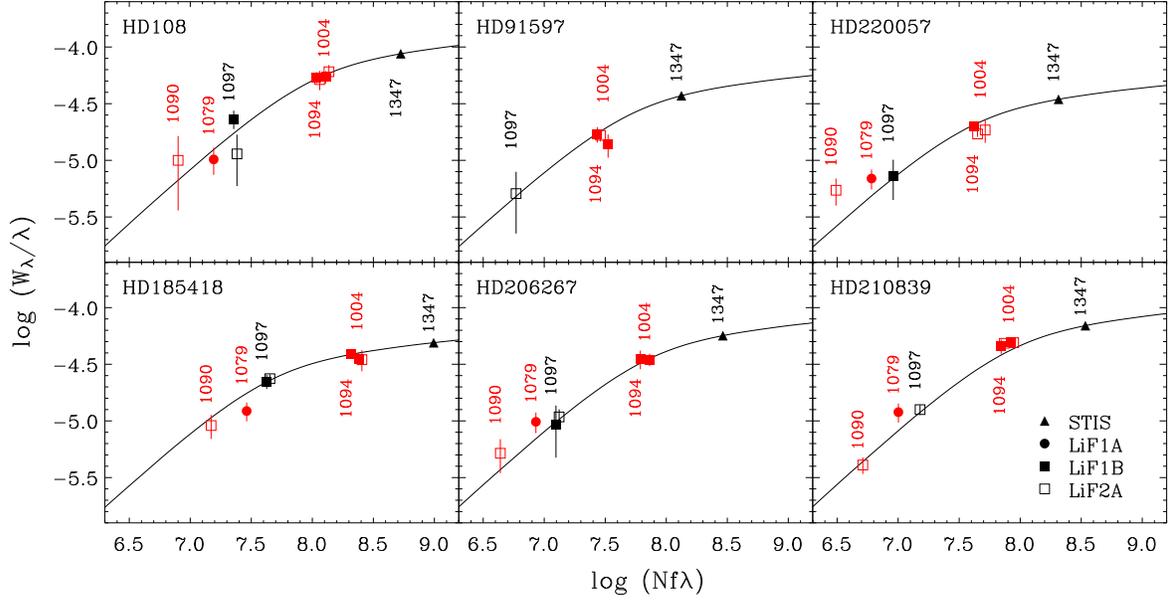}}
\caption{Best fit curve-of-growth with our revised $f$-values (red symbols) for the \ion{Cl}{1} 1004, 1079, 1090 and 1094 \AA\ lines toward our three sample stars HD91597, HD220057 and HD108 (upper panels) and toward three additional archival {\it FUSE} stars (lower panels). For clarity, at each wavelength the data points derived from spectra in each of the two segments have been slightly separated horizontally. \label{fig3}}
\end{figure}

\clearpage

\begin{table}
\caption{\rm{\ion{Cl}{1}} equivalent width measurements in m\AA.\label{tbl-1}}
\vspace{0.2cm}
\begin{tabular}{llllllllll}
\tableline\tableline
Star & 1004.678 & 1079.882 & 1090.739 & 1094.769 & 1097.369 & 1347.240 \\
\tableline
HD91597\tablenotemark{a} & 13.8 $\pm$ 3.2 & $<$ 3.9  & $<$ 3.0  & 18.2 $\pm$ 1.8 & 5.6 $\pm$ 3.1 & 48.1 $\pm$ 2.2\\
HD220057\tablenotemark{a} & 18.6 $\pm$ 4.3 & 6.9 $\pm$ 2.0  & 7.8 $\pm$ 2.1 & 20.0 $\pm$ 1.0 & 8.1 $\pm$ 3.1 & 42.5 $\pm$ 1.7 \\
HD108\tablenotemark{a} & 56.0 $\pm$ 5.4 & 12.6 $\pm$ 3.2  & 6.9 $\pm$ 3.0 & 58.3 $\pm$ 5.7 & 20.5 $\pm$ 3.7 & 117 $\pm$ 3 \\
\tableline
\tableline
HD185418\tablenotemark{b} & 35.7 $\pm$ 3.9 & 12.9 $\pm$ 2.5  & 9.1 $\pm$ 3.5 & 41.2 $\pm$ 4.8 & 25.7 $\pm$ 2.0 & 66.3 $\pm$ 7.0 \\
HD206267\tablenotemark{b} & 33.0 $\pm$ 5.3 & 10.6 $\pm$ 2.2  & 4.8 $\pm$ 1.8 & 38.6 $\pm$ 7.2 & 11.8 $\pm$ 1.9 & 75.3 $\pm$ 7.0\\
HD210839\tablenotemark{b} & 49.8 $\pm$ 3.2 & 11.1 $\pm$ 2.6 & 4.2 $\pm$ 1.1 & 51.8 $\pm$ 2.2 & 12.8 $\pm$ 2.9 & 91.3 $\pm$ 10.0\\
\tableline
\end{tabular}\\
$^{a}${Cycle 4 {\it FUSE} data obtained for the present work. The wavelengths are taken from Morton (2003). When measurements are available in more than one segment, the weighted mean and corresponding 1$\sigma$ formal error are listed.}\\
$^{b}${Archival {\it FUSE} data toward moderately reddened stars (Rachford et al. 2002) used to test the robustness of our results.}
\end{table}

\clearpage

\begin{table}
\begin{center}
\caption{\rm{\ion{Cl}{1}} curve-of-growth results.\label{tbl-2}}
\vspace{0.2cm}
\begin{tabular}{lcccc}
\tableline\tableline
Star & $\log N$(\ion{Cl}{1}) & $b$ & $N$(\ion{Cl}{1})/$N$(\ion{H}{1}) & Refs$^{a}$\\
& (dex) & (km s$^{-1}$) & ($\times$10$^{-8}$) &  \\
\tableline
HD91597 &  13.81$^{+0.23}_{-0.38}$ & 3.93$^{+9.08}_{-0.86}$ & 2.6$\pm$1.8 & 1\\
HD220057 & 14.00$^{+0.19}_{-0.20}$& 3.13$^{+0.37}_{-0.43}$ & 6.8$\pm$3.9 & 2\\
HD108 & 14.41$^{+0.14}_{-0.19}$ & 7.86$^{+1.10}_{-0.60}$ & 7.6$\pm$3.9 & 1\\
\tableline
\tableline
HD185418 & 14.68$^{+0.10}_{-0.06}$ & 3.59$^{+0.14}_{-0.14}$ & 37$\pm$18 & 3\\
HD206267 & 14.15$^{+0.08}_{-0.09}$& 5.32$^{+0.45}_{-0.34}$ & 7.1$\pm$3.2 & 3\\
HD210839 & 14.22$^{+0.05}_{-0.06}$& 6.59$^{+0.36}_{-0.27}$ & 11.7$\pm$3.5 & 3 \\
\tableline
\end{tabular}\\
\end{center}
{$^{a}$ References- 1 $=$ Diplas \& Savage 1994; 2 $=$ Cartledge et al. 2004; 3 $=$ Rachford et al. 2002.}
\end{table}

\clearpage

\begin{table}
\begin{center}
\caption{Revised \rm{\ion{Cl}{1}} $f$-values.\label{tbl-3}}
\begin{tabular}{llll}
\tableline\tableline
$\lambda$  & $f$-values$^a$  & $f$-values$^b$ & Blends  \\
(\AA) & Theory & This work &  \\
 & ($\times$10$^{-3}$)&  ($\times$10$^{-3}$)&  \\
\tableline
1004.678 & 158 & 51.4$^{+5.8}_{-5.1}$ &  \\
1079.882 & 6.98 & 5.60$^{+1.3}_{-1.2}$ &  \\
1090.739 & 6.80 & 2.84$^{+0.4}_{-0.4}$ &  \\
1094.769 & 16.6 & 39.6$^{+4.2}_{-3.8}$ & H$_2$ ($J=$ 6) \\
\tableline
\end{tabular}
\end{center}
\vspace{0.5cm}
$^{a}${Taken from Bi\'emont et al. (1994) and Kurucz \& Peytremann (1975).} \\
$^{b}${Quoted formal errors are 1$\sigma$.}\\
\end{table}

\end{document}